\documentclass[
twocolumn,
english,
aps,
prb,
twocolumn,
superscriptaddress,
amsmath,
amssymb,
longbibliography,
floatfix]{revtex4-2}
\usepackage{graphicx}
\usepackage{epsfig}
\usepackage{color}
\usepackage{amssymb}
\usepackage{amsmath}
\usepackage{array}
\usepackage[loose]{units}
\usepackage{braket}
\usepackage{soul}
\usepackage{hyperref}
\usepackage{cleveref}
\usepackage[caption=false]{subfig}
\usepackage{letltxmacro}
\usepackage[binary-units=true]{siunitx}
\usepackage{dsfont}
\LetLtxMacro{\ORIGselectlanguage}{\selectlanguage}
\makeatletter
\DeclareRobustCommand{\selectlanguage}[1]{%
    \@ifundefined{alias@\string#1}
      {\ORIGselectlanguage{#1}}
      {\begingroup\edef\x{\endgroup
         \noexpand\ORIGselectlanguage{\@nameuse{alias@#1}}}\x}%
}

\usepackage{hyperref}
\definecolor{link_color}{rgb}{1.0,0.0,0.0}
\hypersetup{
pdfborder={0 0 0},
colorlinks=true,
linkcolor=link_color,
urlcolor=link_color,
citecolor=link_color}

\emergencystretch=\maxdimen
\newcommand{\UCBMaterials}{Department of Materials Science and Engineering, University of California Berkeley, Berkeley, CA 94720}
\newcommand{\UCBPhysics}{Department of Physics, University of California at Berkeley, Berkeley, California 94720, United States}
\newcommand{\UCBChem}{Department of Chemistry, University of California at Berkeley, Berkeley, California 94720, United States}
\newcommand{\LBLMSD}{Materials Sciences Division, Lawrence Berkeley National Laboratory, Berkeley, California 94720, United States}
\newcommand{\KavliNano}{Kavli Energy NanoSciences Institute at the University of California at Berkeley, Berkeley, California 94720, United States}
\newcommand{\Foundry}{The Molecular Foundry, Lawrence Berkeley National Laboratory, Berkeley, California 94720, United
States}

\begin{document}



\title{Solving Complex Nanostructures With Ptychographic Atomic Electron Tomography}

\author{Philipp M. Pelz}
\email{philipp.pelz@yahoo.de}
\affiliation{\UCBMaterials}
\affiliation{\Foundry}

\author{Sin\'{e}ad Griffin}
\affiliation{\Foundry}
\affiliation{\LBLMSD}

\author{Scott Stonemeyer}
\affiliation{\LBLMSD}
\affiliation{\KavliNano}
\affiliation{\UCBChem}
\affiliation{\UCBPhysics}

\author{Derek Popple}
\affiliation{\LBLMSD}
\affiliation{\KavliNano}
\affiliation{\UCBChem}
\affiliation{\UCBPhysics}

\author{Hannah Devyldere}
\affiliation{\UCBMaterials}

\author{Peter Ercius}
\affiliation{\Foundry}

\author{Alex Zettl}
\affiliation{\UCBMaterials}
\affiliation{\LBLMSD}
\affiliation{\KavliNano}
\affiliation{\UCBPhysics}

\author{Mary C. Scott}
\affiliation{\UCBMaterials}
\affiliation{\Foundry}

\author{Colin Ophus}
\email{cophus@gmail.com}
\affiliation{\Foundry}

\date{\today}
\begin{abstract} 

Transmission electron microscopy (TEM) is a potent technique for the determination of three-dimensional atomic scale structure of samples in structural biology and materials science. In structural biology, three-dimensional structures of proteins are routinely determined using phase-contrast single-particle cryo-electron microscopy from thousands of identical proteins, and reconstructions have reached atomic resolution for specific proteins.  In materials science, three-dimensional atomic structures of complex nanomaterials have been determined using a combination of annular dark field (ADF) scanning transmission electron microscopic (STEM) tomography and subpixel localization of atomic peaks, in a method termed atomic electron tomography (AET). However, neither of these methods can determine the three-dimensional atomic structure of heterogeneous nanomaterials containing light elements. Here, we perform mixed-state electron ptychography from 34.5 million diffraction patterns to reconstruct a high-resolution tilt series of a double wall-carbon nanotube (DW-CNT), encapsulating a complex $\mathrm{ZrTe}$ sandwich structure. Class averaging of the resulting reconstructions and subpixel localization of the atomic peaks in the reconstructed volume reveals the complex three-dimensional atomic structure of the core-shell heterostructure with \SI{17}{\pico\meter} precision. From these measurements, we solve the full $\mathrm{Zr_{11}Te_{50}}$ structure, which contains a previously unobserved $\mathrm{ZrTe_{2}}$ phase in the core. The experimental realization of ptychographic atomic electron tomography (PAET) will allow for structural determination of a wide range of nanomaterials which are beam-sensitive or contain light elements.

\end{abstract}


\maketitle




Knowledge of the three-dimensional atomic structure of natural and manufactured materials allows us to calculate their physical properties and deduce their function from first principles. Because of this, methods for atomic structure determination have been a key area of research in the biological and physical sciences, including x-ray crystallography \cite{blakeley2015sub} and nuclear magnetic resonance spectroscopy \cite{jackman2013application}. For crystalline samples, micro-crystal electron diffraction provides data of similar quality to X-ray diffraction for solving structures \cite{nannenga2019cryo}. More recently, cryo-electron microscopy has become the dominant method for atomic structure determination of molecules, either from ensembles fulfilling the single-particle assumption \cite{cheng2018single}, or by cryo-electron tomography and subsequent subtomogram averaging \cite{himes2018emclarity}. However, these methods require averaging of many near-identical structures. One method which is capable of solving structurally and chemically heterogeneous nanostructures is atomic electron tomography (AET) using scanning transmission electron microscopy \cite{AET_chen2013Wtip, AET_goris2013elemental, AET_xu2015three, AET_haberfehlner2015formation, AET_yang2017deciphering, AET_zhou2019observing}. However, the dark field imaging method used in these AET studies produces very little contrast for light elements, and requires too much electron dose to be used for beam-sensitive samples. Phase contrast imaging such as the method used in cryo-electron microscopy overcomes these obstacles, and has recently been extended to include multiple scattering \cite{tem_tomo_ren2020_algo} and demonstrated experimentally at sub-nm resolution \cite{tem_tomo_whittaker2020_clay}. A STEM-based phase contrast method which is capable of removing residual aberrations of the probe and enhancing the resolution is ptychography \cite{Hegerl_Hoppe_1970,nellist1995resolution,Yang_2016}, recently demonstrated with deep sub-\AA{}ngstrom resolution \cite{Jiang_2018, ptycho_chen2021latticevibrations}. Here, we use ptychographic atomic electron tomography (PAET) to solve the 3D atomic structure of a zig-zag double wall-carbon nanotube (DW-CNT) encasing a structurally complex Zr-Te shell and a previously unseen Zr-Te core structure.


\begin{figure*}[ht!]
    \includegraphics[width=\textwidth]{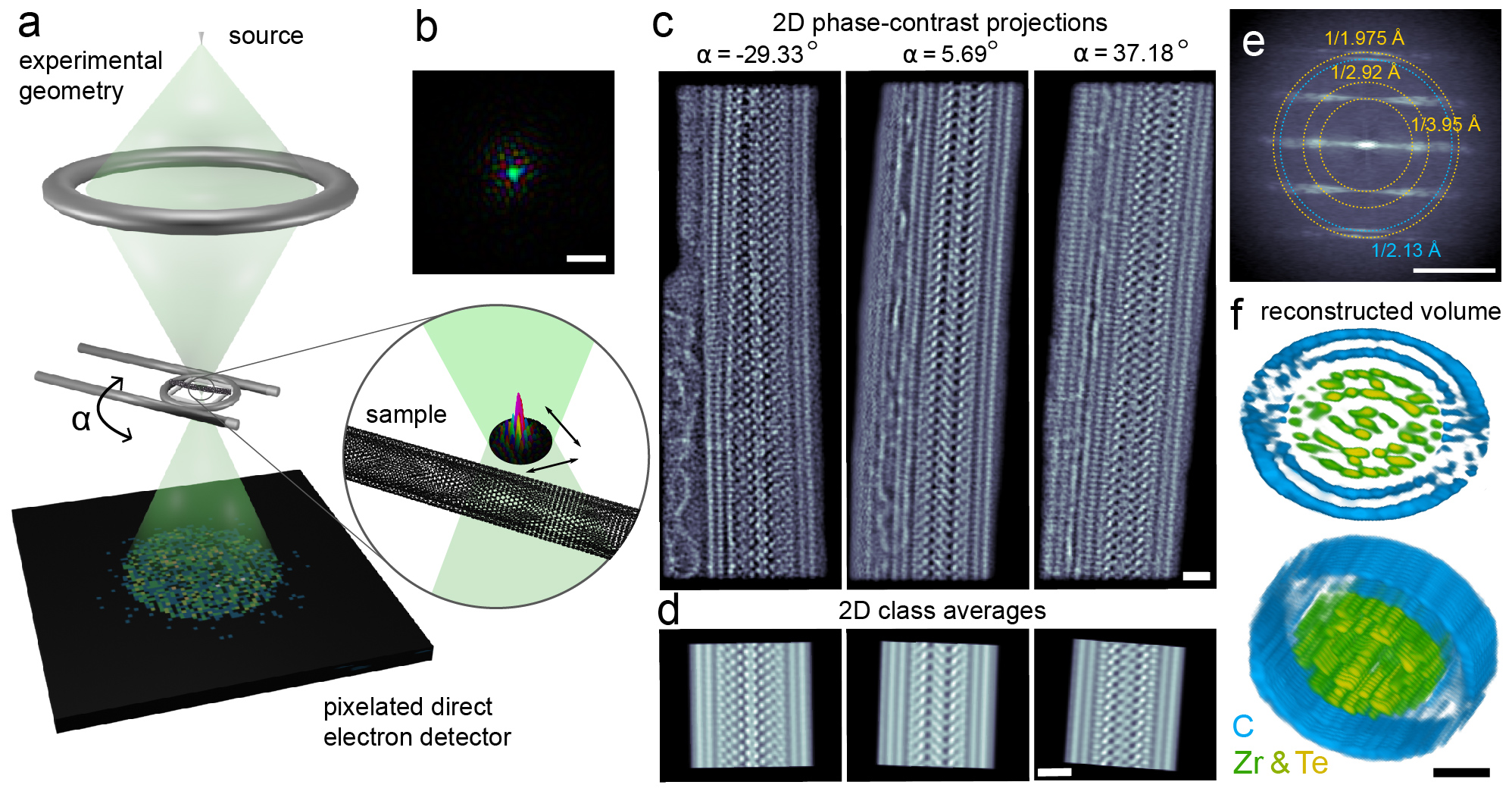}
    \caption{\label{fig:experimental_setup} \textbf{Ptychographic atomic  electron tomography (PAET) of a complex nanostructure.} (a) A converged electron beam is scanned over a nanoscale sample which is tilted around the $\mathrm{\alpha}$ axis of a high-precision tomography holder. Four-dimensional scanning diffraction datasets are recorded at every tilt angle using a direct electron detector. (b) Reconstructed complex wavefunction of an example STEM probe. (c) Examples of phase-contrast projection images, reconstructed using mixed-state electron ptychography. (d) High-SNR class averages calculated from the images in (c). (e) Average diffraction pattern of all 36 tilt angles with some scattering angles labeled. (f) Two orientations of the 3D reconstructed volume from the 2D unit cell averages in (d). All real space scales bars are \SI{10}{\angstrom}, scale bar in (e) is  \SI{0.5}{\angstrom^{-1}}.}
\end{figure*}


Low-dimensional van der Waals (vdW) materials such as the transition metal di- and tri-chalcogenide families or CNTs exhibit a range of desirable properties that often emerge as the material thickness reaches the one- or two-dimensional thickness limit \cite{encaps_philp2003encapsulated, encaps_pham2020emergence,encaps_pham2018torsional}. In order to synthesize otherwise unstable vdW materials, encapsulation inside CNTs has been developed as a strategy to stabilize quasi one-dimensional structures \cite{encaps_meyer2019metal,encaps_kanda2020efficient,  encaps_kashtiban2021linear} of transition metals. The encapsulation approach protects the interior structure from oxidation, and can result in chiral structures \cite{encaps_pham2018torsional, encaps_kanda2020efficient} or fillings with high aspect ratio \cite{encaps_cain2021ultranarrow} that exhibit properties differing drastically from their bulk counterparts. While electron microscopy has played an important role in the identification and characterization of the encapsulated phases, the detailed 3D structure of the interior nanowire is not always clear. For example, the encased material's structure often exhibits a dependence on the diameter of the nanotube \cite{encaps_meyer2019metal}, and can form complex three-dimensional structures inside the CNT, such as core-shell structures \cite{encaps_pelz2021phase} and multilayer moir\'{e} structures \cite{encaps_stonemeyer2022targeting}. In the latter cases, the atomic structure cannot be uniquely determined from a single projection, and three-dimensional imaging is paramount for structure-function determination.

The experimental setup for PAET in the STEM is shown in Fig.~\ref{fig:experimental_setup}a. A converged electron probe is raster-scanned over the sample, with one diffraction pattern recorded at every probe position using a fast-framing direct-electron detector operating at \SI{87}{\kilo\hertz}.  The nanotube is tilted around its axis using a specialized dual-axis tomography holder, and at each tilt angle the scanning diffraction measurements are repeated. We reconstruct the partially coherent illumination wave function from each four-dimensional dataset, where the dominant mode of an example probe's complex wavefunction is shown in Fig.~\ref{fig:experimental_setup}b. The probe wavefunctions and their positions are reconstructed jointly with the sample object wave. Fig.~\ref{fig:experimental_setup}c shows the phase of the object wave from three of the experimental projections.  The strong contrast of the carbon atoms in the 2D projections allows us to determine a zig-zag nanotube configuration and the dimensions of the semi-minor and semi-major axes of the elliptical DW-CNT (see Methods Section \ref{sec:atomic_model}), information unobtainable from the ADF signal at the same electron dose.

\begin{figure*}[ht!]
    \includegraphics[width=6.0in]{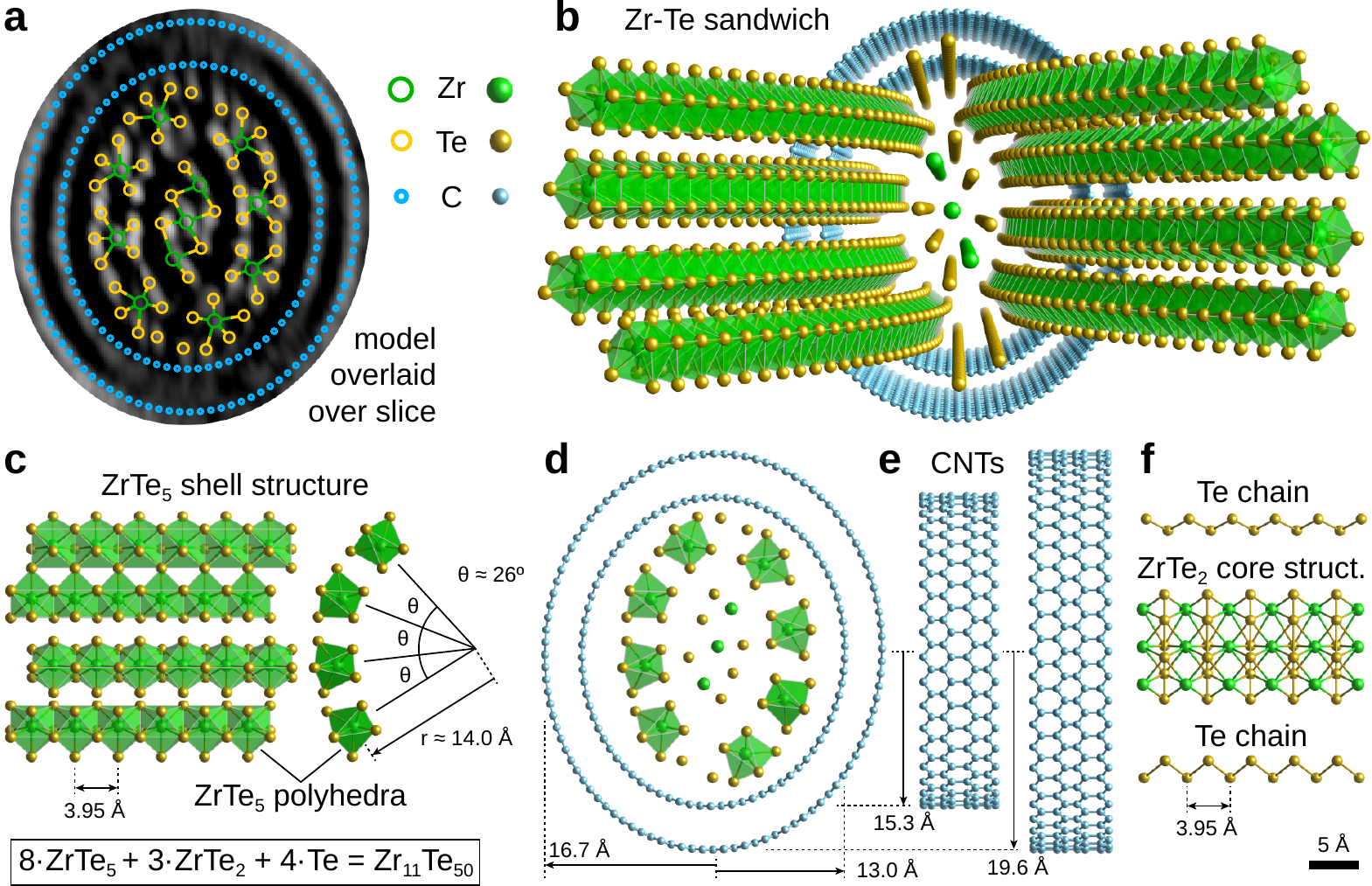}
    \caption{\label{fig:ZrTeModel} \textbf{DW-CNT-encapsulated $\mathrm{Zr_{11}Te_{50}}$ sandwich structure.} (a) atomic model overlaid over vertical slice through the reconstructed volume. (b) Three-dimensional model of the $\mathrm{Zr_{11}Te_{50}}$ structure with folded-out one-dimensional $\mathrm{ZrTe_{5}}$ chains. (c) Left panel: side view of four $\mathrm{ZrTe_{5}}$ chains building one side of the sandwich structure. Right panel: Front view of four $\mathrm{ZrTe_{5}}$ chains of one side of the sandwich structure. (d) Front view of the full atomic model showing the elliptical CNTs and the spatial extents of their semi-minor and semi-major axes. (e) Side view of the DW-CNT with zig-zag configuration. (f) Side view of the $\mathrm{ZrTe_{2}}$ core structure and the $\mathrm{Te}$ chains.}
\end{figure*}

To increase the signal-to-noise ratio of the phase-contrast projections, and enable determination of the 3D structure of the center of the tube even with a limited experimental tilt range, we computed class averages along the nanotube, with 3 classes shown in Fig.~\ref{fig:experimental_setup}d which correspond to the images in Fig.~\ref{fig:experimental_setup}c. The size of these class images were chosen to include 11 repeats of the core structure, and approximately 23 repeats of the DW-CNT structure. This corresponds to a rational approximate of the irrational ratio of the repeat length for these two incommensurate structures. Fig.~\ref{fig:experimental_setup}e shows the average diffraction pattern of the 38 phase-contrast projections, which displays a high degree of periodicity along the tube. The \SI{1/2.13}{\angstrom^{-1}} reflection of the graphene lattice is clearly visible, as well as the 1st and 2nd diffraction order of the Zr-Zr and Te-Te lattice spacing of \SI{1/3.95}{\angstrom^{-1}} and \SI{1/1.975}{\angstrom^{-1}}. The unit-cell averaging procedure yielded a high-SNR image of the core structure, while the incommensurate \SI{2.13}{\angstrom} spacing of the graphene lattice in the carbon nanotube is highly suppressed. 


We have reconstructed the 3D volume shown in Fig.~\ref{fig:experimental_setup}f, from the 2D class images in shown in Fig.~\ref{fig:experimental_setup}d, using the methods  described in Methods Sec.~\ref{subsec:tomo}. From this reconstruction, we can clearly identify the outer 2 shells which correspond to the DW-CNT, which have been colored in blue. We can also see the complex Zr-Te interior structure, colored in green and gold. This structure is highly periodic along the tube axis, but does not possess a periodic crystal structure in the plane perpendicular to the nanotube axis. Instead, the Zr-Te structure possesses a distinct core and shell structure, with a high degree of variation in the structural sub-units making up each. Most atoms in the Zr-Te core and shell can be directly resolved, while some atoms close to the edges along the missing wedge direction appear as elongated columns. From the reconstructed volume, we have measured the 3D positions of 578 Zr and Te atoms to subpixel precision using the procedures developed in AET routines \cite{AET_chen2013Wtip, tem_tomo_ren2020_algo}. The chemical species of the Te and Zr atoms were determined by analyzing the number of nearest-neighbors and the local coordination (Methods section \ref{sec:atomic_model}).

\begin{figure*}[ht!]
    \includegraphics[width=6.0in]{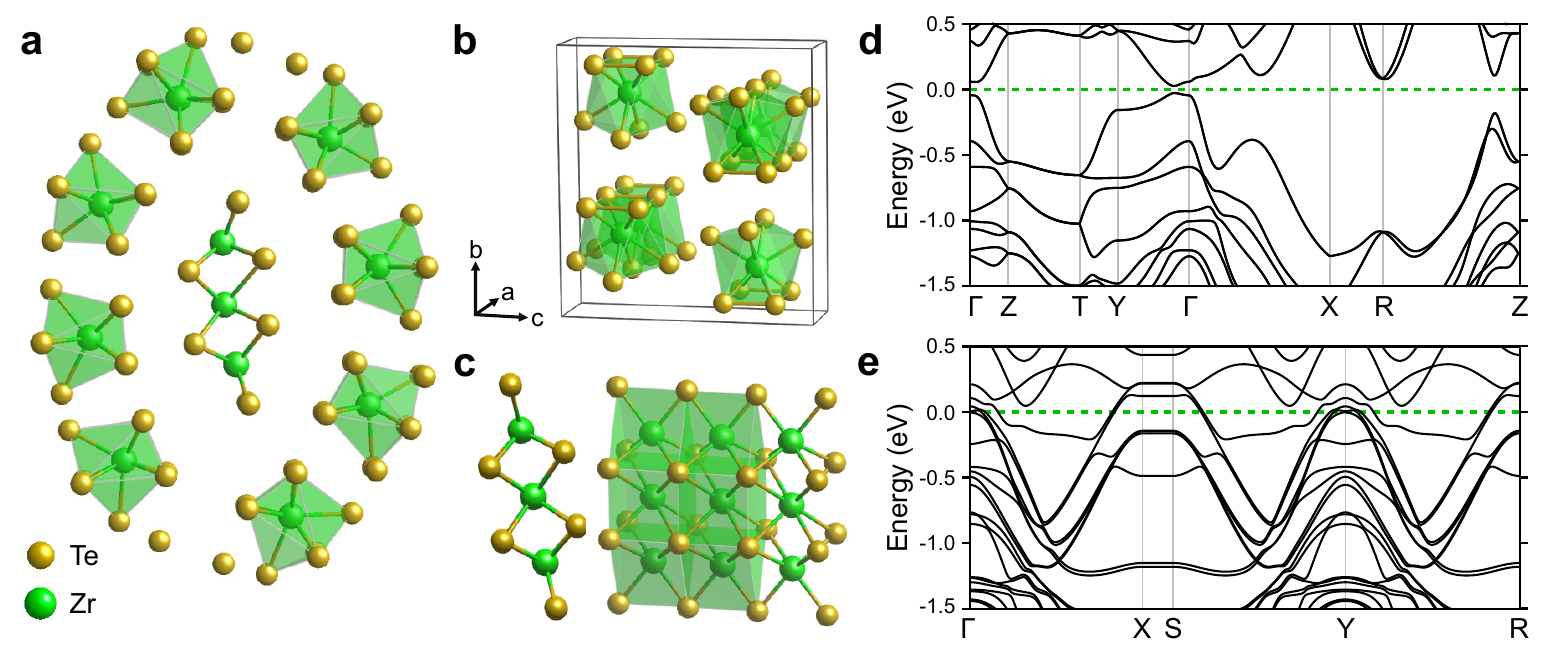}
    \caption{\label{fig:theory} \textbf{First principles calculations of Zr$_{11}$Te$_{50}$}. (a) DFT optimized Zr$_{11}$Te$_{50}$ structure. (b) Crystal structure of $Cmcm$ ZrTe$_5$ (c) Fully optimized inner ZrTe$_2$ structural unit d) Calculated electronic band structure of $Cmcm$ ZrTe$_5$. e) Calculated electronic band structure of inner ZrTe$_2$ structural unit. For both band structures, the Fermi level is set to 0 eV and is marked by the green dashed line.}
\end{figure*}

The configuration and elliptical dimensions of the DW-CNT were determined by first excluding chiral CNTs based on the Fourier spectrum of the projections shown in Fig.~\ref{fig:experimental_setup}e. The armchair configuration was excluded based on the observation of a strong reflection at \SI{1/2.13}{\angstrom^{-1}} in the direction of the nanotube, which is present only in zig-zag nanotubes where the \SI{2.13}{\angstrom} spacing is oriented across the nanotubes. The dimensions of the CNTs were determined by maximizing the agreement of the atomic coordinates with the reconstructed projected intensity of the CNTs from the volume in Fig.~\ref{fig:experimental_setup}f. We determined best-fit chiral indices of $\mathrm{(50,0)}$ and $\mathrm{(39,0)}$ for the outer and inner CNT.


Fig.~\ref{fig:ZrTeModel}a shows the full atomic model overlaid over a 2D slice through the reconstructed phase-contrast volume. Fig.~\ref{fig:ZrTeModel}b shows a three-dimensional rendering of the structure, with the outermost Zr-Te shells peeled back to display the local coordination. Inside the DW-CNT, eight one-dimensional $\mathrm{ZrTe_{5}}$ clusters encapsulate a three unit-cell wide $\mathrm{ZrTe_{2}}$ structure. The outer $\mathrm{ZrTe_{5}}$ clusters are split into two groups of 4 $\mathrm{ZrTe_{5}}$ sub-units, shown in Fig.~\ref{fig:ZrTeModel}c. Each of these clusters has a 1D chain structure along the nanotube axis, with the same orientations \cite{furuseth1973crystal}. The 4 groups are circumscribed inside the inner CNT wall, with an angle of approximately $26^\circ$ between each cluster and a radius of \SI{14}{\angstrom} to the central Zr atom. The overall geometric arrangement of the full structure is shown in Fig.~\ref{fig:ZrTeModel}d and e, with the DW-CNT major and minor axes lengths labeled.

On the top and bottom of the encapsulated Zr-Te structure, there are two single-atom Te chains. One-dimensional Te-Te chains have been observed before as stable structures encapsulated in CNTs \cite{qin2020raman}. The relative position of the Te-Te chains and the $\mathrm{ZrTe_{2}}$ core structure are shown in Fig.~\ref{fig:ZrTeModel}f. The $\mathrm{ZrTe_{2}}$ is a previously unobserved phase of Zr-Te, which we examine in more detail in the following section. The overall structure has a stoichiometry of $\mathrm{Zr_{11}Te_{50}}$, and a striking elliptical structure which may be caused by the intercalation of the ZrTe structures into the DW-CNT.






To investigate the stability and electronic structure of our proposed $\mathrm{Zr_{11}Te_{50}}$ structure, we have performed first principles calculations, as described in Methods Sec.~\ref{sec:dft}. The resulting optimized atomic structure is depicted in Fig.~\ref{fig:theory}a. As previously discussed, the outer tube is well described as one-dimensional chains of ZrTe$_5$ which consist of face-sharing Zr-Te polyhedra where each Zr is 8-coordinated with Te. These 8-coordinated Zr-Te chains are distorted versions of those that make up bulk $Cmcm$ ZrTe$_5$ crystal where each Zr has three pairs of identical Zr-Te bond lengths, with the two remaining Zr-Te bonds being different. The coordination of the outer ring of Zr in our Zr$_{11}$Te$_{50}$ is structurally the same with slight deviations from the ``paired'' Zr-Te bond lengths (e.g. 2.955 \AA{} -vs- 2.960 \AA{}). We present the calculated electronic band structure of the bulk ZrTe$_5$ from which our outer structure is derived in Fig. \ref{fig:theory}d -- we confirm the previously-reported topological insulating phase with the band inversion near to $\Gamma$ \cite{Manzoni_et_al:2016}.

The inner core ZrTe$_2$ structure presents an intriguing case for structural analysics. We find two different coordinations for Zr in the innermost layer. The central Zr is 8-fold coordinated with Te, with the other two Zr sites being 6-fold coordinated with Te. We next isolated this ZrTe$_2$ structural unit and performed a full structural optimization to remove the influence of any confinement/pressure from the sandwich structure, with the result depicted in Fig.~\ref{fig:theory}c. This new ZrTe$_2$ structure adopts the space group \textit{Pmmm}, and forms a thin two-dimensional slab. In fact, the ground state polymorph, $1T$-ZrTe$_2$ (space group $P\overline{3}m1$), has only recently been grown by molecular beam epitaxy \cite{Tsipas_et_al:2018}. This $1T$ phase is 6-fold coordinated and forms a van-der-Waals two-dimensional structure with measurements and theory suggesting it is a Dirac semimetal \cite{Tsipas_et_al:2018}. The innermost Zr coordination is more unusual. While this Zr site is 8-fold coordinated as in the ZrTe$_5$ chains, it now forms a regular octahedron where each Zr-Te bond length is the same. The calculated electronic band structure of this new structural phase is shown in Fig.~\ref{fig:theory}e. Our $Pmmm$ ZrTe$_2$ phase is a metal with primarily hole pockets at the Fermi level -- the low-dimensional nature of the structure is also confirmed with the flat bands in the X to S directions (short axis). Finally, using symmetry indicators we predict this new phase to be a Dirac semimetal, akin to its stable $P\overline{3}m1$ polymorph \cite{Tsipas_et_al:2018}.




In summary, we have experimentally demonstrated atomic structure determination of a complex nanomaterial by phase-contrast ptychographic atomic electron tomography. In addition to showing contrast improvement compared to AET for the weakly-scattering carbon atoms, PAET simultaneously recovers partial coherence effects, probe positions and probe aberrations present in the experiment. This in turn allows focusing at lower resolution and an overall lower dose due to less pre-exposure irradiation of the sample. Further improvements of the experimental protocol will allow skipping the unit-cell averaging step in the reconstruction and enable imaging of single light atoms at atomic resolution.  We have used PAET to solve the structure of a double-walled carbon nanotube encapsulating a complex Zr$_{11}$Te$_{50}$ nanowire structure, which contains both previously observed linear chain structures of ZrTe$_5$ and Te-Te, and a previously unobserved ZrTe$_2$ structure. Density functional calculations both confirm the stability of our Zr-Te model structure, and elucidate its electronic properties. We expect that PAET will find widespread application for solving the structures of complex materials which contain light elements, weakly-scattering structures, and beam-sensitive materials.

\medskip

\section*{Methods}

\subsection{Sample preparation}
Encapsulated $\mathrm{Zr_{11}Te_{50}}$ species are synthesized within CNTs using protocols similar to those for the growth of confined $\mathrm{TaTe_2}$, $\mathrm{NbSe_3}$, and $\mathrm{HfTe_3}$ structures \cite{encaps_stonemeyer2022targeting, encaps_pham2018torsional,encaps_meyer2019metal}. The $\mathrm{Zr_{11}Te_{50}}$ species can be synthesized following the synthesis of $\mathrm{ZrTe_3}$, where stoichiometric quantities of Zr powder and Te shot (\SI{~450}{\milli\gram} total), \SIrange{1}{2}{\milli\gram} of end-opened multi-walled CNTs with an inner diameter ranging from \SI{1.0}{\nano\meter} to \SI{10.0}{\nano\meter} (CheapTubes, 90\% SW-DW CNT), and \SI{5}{\milli\gram\per\centi\meter^3}(ampoule volume) of $\mathrm{I_2}$ are sealed under vacuum (\SI{1.33E-6}{\milli\bar}) in a quartz ampoule. The ampoule is heated in a single-zone furnace at \SI{550}{\degree} for 3-5 days, cooled to \SI{350}{\degree} over 3 days, then cooled to room temperature over 1-2 days. The Zr-Te filled CNTs are dispersed in isopropyl alcohol by bath sonication for 1 hour and drop-cast onto lacey carbon TEM grids for subsequent electron microscopy analysis.

\subsection{Data Acquisition}

A tomographic tilt series was acquired from a $\mathrm{Zr_{11}Te_{50}}$ DW-CNT using the TEAM 0.5 microscope and TEAM stage \cite{Ercius_Boese_Duden_Dahmen_2012} at the National Center for Electron Microscopy in the Molecular Foundry. 
Before the tilt series, the sample was beam showered for 30 minutes. We recorded four dimensional-scanning transmission electron microscopy (4D-STEM) datasets \cite{ophus2019four} with full diffraction patterns over\num{1600x600} probe positions at each tilt angle. The diffraction pattern images were acquired with the 4D Camera prototype, in-house developed in collaboration with Gatan Inc., a direct electron detector with \num{576x576} pixels and a frame rate of \SI{87}{\kilo\hertz} \cite{ercius20204d}, at 80 kV in STEM mode with a \SI{25}{\milli\radian} convergence semi-angle, a beam current of \SI{40}{\pico\ampere}, estimated from 4D camera counts, a real-space pixel size of \SI{0.4}{\angstrom}, and camera reciprocal space sampling of \SI{173.6}{\micro\radian} per pixel. These settings amounted to an accumulated fluence of \SI{1.79e4}{\elementarycharge\per\angstrom^2} per projection and \SI{6.28e5}{\elementarycharge\per\angstrom^2} for the whole tilt series. The tilt series was collected at 38 angles with a tilt range of +\num{63} to \num{-58} degrees. To minimize the total electron exposure, focusing was performed at a resolution of 80 kX before switching to high magnification for data collection.

\subsection{Ptychographic reconstruction}

The raw 4D-STEM datasets of size \SI{650}{\giga\byte} per tilt angle and \SI{22.75}{\tera\byte} in total were electron-counted using the open-source stempy software \cite{stempy} on the Cori supercomputer at NERSC and saved in a sparse linear-index encoded electron event representation (EER), using \SI{~6.5}{\giga\byte}  storage per tilt. Crop-outs containing only the scan positions covering the $\mathrm{Zr_{11}Te_{50}}$ DW-CNT for all tilts were determined from the manually aligned annular bright-field images of the tilt series for ptychographic reconstruction. 
Initial guesses for the defocus aberration for each tilt were manually obtained with an interactive real-time implementation \cite{Pelz_2021} of the single-sideband ptychography (SSB) method \cite{pennycook2015efficient}. The data in EER format was further preprocessed by centering the bright-field disk, symmetric cropping to a maximum detector angle of 2$\alpha$, with $\alpha$ the semi-convergence angle, and binning to a detector size of \num{88x88} pixels. The final cropped and preprocessed data of the whole tilt series, compressed with the gzip compression algorithm, had a total size of \SI{12.3}{\giga\byte}.
Phase-contrast images, probe positions and a low-rank approximation of the partially coherent illumination were jointly reconstructed using 115 iterations of the Least-Squares Maximum Likelihood (LSQML) method with gradient-based scan-position correction \cite{LSQML, Chen_2020}, with the parameters in Suppl. Table II. The maximum thickness for a sample to fulfill the projection approximation of ptychography for a numerical aperture of \SI{25}{\milli\radian} is \SI{9}{\nano\meter}. The largest dimension of the DW-CNT along the beam direction is \SI{3.9}{\nano\meter}, it is therefore within the limit of the projection approximation for single-slice ptychography. Due to the slight tilt of the nanotube, the projected thickness reaches \SI{6}{\nano\meter} for some projections. For the best reconstruction quality of projections where the nanotube is tilted along the beam direction, we reconstruct slices along the nanotube separately with different probes. 

\subsection{Preprocessing for phase-contrast tomography}
\label{subsec:preoricessing}
Since low spatial frequencies are weakly transferred in noisy 4D-STEM datasets collected close to the focus condition \cite{ptycho_yang2016midiptycho}, the ptychographic tilt-series reconstructions display characteristic halos around the nanotube. With the prior knowledge that the nanotubes are suspended over vacuum, we manually create vacuum masks around the nanowire and set the vacuum phase equal to the lowest phase value of the halo close to the nanotube. We then performed a first manual alignment using the Tomviz tomography software \cite{tomviz}. Subsequently we scaled all projections with a scalar value to have the same integrated phase shift along the direction perpendicular to the tilt axis. Tomviz was also used for visualization of the reconstructed volume.

\subsection{Tomographic reconstruction}
\label{subsec:tomo}
From the initially-aligned tilt series a 3D reconstruction was performed using the fast adaptive shrinkage-thresholding algorithm \cite{Goldstein_Studer_Baraniuk_2014}, an accelerated gradient algorithm with adaptive stepsize for faster convergence. To compute the forward- and backward projections, we used the generalized ray transform interface of the Operator Discretization Library \cite{ODL_2018} in a 3D parallel-beam Euler geometry with an GPU-accelerated backend of the ASTRA tomography toolbox \cite{ASTRA_2015}. To increase the accuracy of the projections, we used the trilinear interpolation feature of the ASTRA library to compute the forward and inverse ray-transforms. To minimize the translational and angular misalignments, we adopt the following procedure. We perform a grid search for a range of global tilt axis rotation errors in the range of -2 to 6 degrees. For each of these global rotation axis guesses, we perform a local refinement by an iterative projection matching approach \cite{tomo_dengler1989multiresolution} with simulated annealing. In this approach, all three Euler angles are varied by a randomly picked value in the range of \num{-1} to \num{+1} degrees and the calculated projection error compared with the current projection error after a full reconstruction. The lowest-error angles are then used as new initial angles for the next tomographic reconstruction. Initially the reconstruction was performed on the full 2D projections. After a reconstruction, the auto-aligned projections were checked for unresolved unit-cell hops. These unit-cell hops in the tilt series were fixed manually in tomviz \cite{tomviz}, and then another round of algorithmic auto-alignment was performed. This process of manual alignment and auto-alignment was repeated for 16 rounds until no further improvements could be made. The 2D class averaged projections were then manually aligned to the final aligned full projections to avoid unit-cell hops. Using this approach, the reconstruction with the 2D class averages converged to a minimum R-factor of \SI{5.4}{\percent} for a global tilt axis rotation of 6.3 degrees. The reconstructed volume from the full 2D projections is \SI{27.7 x 11.6 x 11.6}{\nano\meter} in size, while the volume reconstructed from 2D class averages is \SI{5.88 x 5.3 x 5.3}{\nano\meter} in size.

\subsection{Atom tracing}
\label{sec:atom_tracing}
The 3D atomic positions of the Zr and Te atoms were determined using the following procedure developed in \cite{tem_tomo_ren2020_algo}. (I) all local intensity maxima were identified from the 3D reconstruction and added to a candidate list. From the initial candidate list, peaks which were within a minimum distance of \SI{2}{\angstrom} of a higher-intensity peak were deleted. (II) The initial list of peak positions was refined by fitting a 3D Gaussian function to each peak after subtracting neighboring peaks within a maximum radius of \SI{4.5}{\angstrom}. Using this initial atom candidate list, we added, refined, and merged new unidentified peaks for 4 iterations in the following order:
(III) Subtract the fitted Gaussians of all current peak candidates from the reconstruction volume. (IV) Add new candidate peaks over an intensity threshold of \num{50} to the candidate list. (V) Refine the positions of all atom candidates as in (II) for 4 iterations. (VI) Merge peaks that are closer than a minimum distance of \SI{2.2}{\angstrom} after position refinement. (VII) Refine the positions of all atom candidates as in (II) for 4 iterations. (VIII) go to (III) if iterations not done. (IX) A final set of 4 positions refinement iterations as in step (II) yielded the set of coordinates of 714 candidate atoms. The rotation of the nanotube with respect to the beam propagation normal plane was determined as \SI{5.2}{\degree} and the candidate atoms were then rotated onto this normal plane. Atomic columns along the tube were identified by projecting the volume along the tube and 2D peak finding. Atom candidates were that were farther than \SI{2}{\angstrom} from the next column were snapped onto the column in the projection along the tube, and duplicate atoms were removed. This procedure resulted in 578 candidate atoms. 

\subsection{Atomic model construction}
\label{sec:atomic_model}

From the candidate atoms, we have constructed the Zr-Te model structure. We note that for most sites, it was not possible to directly determine the atomic species from the reconstructed 3D phase-contrast signal at the experimental signal-to-noise ratio. The prototype 4D camera detector is mounted off-center relative to the ADF detectors, such that the simultaneous ADF signal could not be used for tomographic reconstruction. We expect better chemical contrast in future experiments with both ADF and 4D-STEM signals \cite{chang2020ptychographic, Yang_2016}. Chemical species of all but 2 atomic columns in the nanotube could be determined from the coordination environment with the following procedure. For each atom candidate, we determined the number of nearest neighbors (NNs) in a shell of \SI{3.1}{\angstrom} radius. As candidate stable structures we considered all known stable Zr and Te containing structures, including ZrTe$_2$, ZrTe$_3$ and ZrTe$_5$. In those structures, the Zr atoms are bonded to six or eight Te atoms, while the Te atoms can be bonded in different geometries to 2-4 Zr atoms. We therefore created two classes for atoms with less than four neighbors and those with more than four neighbors. The result of this classification is shown in supplementary Fig 7. We then assigned the columns with majority of atoms with more than 4 NNs as Zr atoms and the columns with majority of atoms in the class with less than 4 NNs as Te atoms.

This procedure left two atomic columns in the center of the nanotube that could not be assigned based on NNs. These are marked with circles in supplementary Fig 7. Those two ambiguous site identities were tested using DFT calculations, where in all cases we used the lower energy stoichiometries.
From the chemical classification based on NNs and the candidate atomic positions, the 8 outermost ZrTe$_5$ units and their positions / angles were identified. Next, we identified two isolated chains of atoms following a zigzag pattern at the top and bottom of the tube as positioned in Fig.~\ref{fig:ZrTeModel}. These sites were assumed to be composed of Te atoms, due to previous observations of Te chains encapsulated in carbon nanotubes \cite{qin2020raman}, and the presence of weakly-bound Te chains in the ZrTe$_5$ structural units. Finally, the core was determined to be composed of a central Zr atom with 8 Te neighbors, and two partially-coordinated Zr atoms, forming a small 2D ribbon of ZrTe$_2$ with a previously unreported structure.

The DW-CNT configuration was determined in the following way. CNTs are characterized by their chiral indices $\mathrm{(n,m)}$. Chiral nanotubes can be excluded for our sample because the diffraction pattern of chiral nanotubes follows a Bessel function radial symmetry \cite{Cochran_Crick_Vand_1952} with oscillating layer lines. In the diffraction patterns of the experimental projections in supplementary Fig. 1 (insets) it can be seen that the \SI{1/2.13}{\angstrom^{-1}} graphene reflection is not an oscillating layer line.  This leaves armchair configuration $\mathrm{(k,k)}$ or zig-zag configuration $\mathrm{(k,0)}$ nanotubes as possible options. In armchair configuration nanotubes, the \SI{2.13}{\angstrom} spacing lies across the nanotube and can be suppressed by material in the tube, while in zig-zag configuration the \SI{2.13}{\angstrom} spacing lies along the tube and is strongly visible at the edges of the tube where it is overlapped by no other material. This is the case in our sample, and we therefore exclude the armchair configuration, which leaves us to determine the chiral indices $\mathrm{k_1}$ and $\mathrm{k_2}$ of the outer and inner nanotube, and the determination of the semi-minor and semi-major axis dimensions of the nanotube. We determined the chiral indices by maximizing the the intensity of the pixels sampled by CNT atomic coordinates overlaid over the 2D projection of the reconstructed volume along the tube, as shown in supplementary Fig 7. The sampled intensity was maximized for chiral numbers $\mathrm{k_1 = 50}$ and $\mathrm{k_2 = 39}$, with semi-minor axes of \SI{16.7}{\angstrom} and \SI{13.0}{\angstrom} and semi-major axes of \SI{19.6}{\angstrom} and \SI{15.3}{\angstrom}. The only parameter that could not be determined uniquely was the relative rotation of the CNTs to each other, as the 2D unit-cell averaging procedure reduced the atomic contrast for the CNT. As such, the relative rotation of the CNTs shown in Fig.~\ref{fig:ZrTeModel} is only one possible configuration.

\subsection{Estimation of Precision}
\label{sec:precision}

To calculate the precision of our position measurements, we first investigated if the nanotube can be modeled with a linear model or if a full multislice treatment is necessary. To this end, we simulated one projection both with a linear forward model as in \cite{AET_yang2017deciphering} and with a multislice simulation incorporating all partial spatial and temporal coherence effects, followed by mixed-state ptychographic reconstruction and unit cell averaging. 
First, we created 12 different atomic models where the Zr and Te atoms and atom positions are identical, but the carbon nanotube is displaced by \SI{2.13/12}{\angstrom} every time, such that the averaging removes atomic contrast of the nanotube. We then fitted a linear model to this average structure by determining the H- and B-factors that minimize the R-factor between model and experimental reconstruction. This linear model is shown in the middle panel of supplementary Fig. 4 and achieves and \mbox{R-Factor} of \SI{12}{\percent}. Then, we performed multislice 4D-STEM simulations using the PRISM algorithm \cite{ophus2017fast} in the abTEM simulation code \cite{madsen2021abtem}, incorporating partial spatial coherence with a FWHM source size of \SI{80}{\pico\meter} and partial temporal coherence with a chromatic defocus spread of \SI{6}{\nano\meter}, as calculated from the chromatic aberration coefficient and energy spread of the TEAM 0.5 microscope at \SI{80}{\kilo\eV}, and with the experimental dose of \SI{1.72e4}{\elementarycharge\per\angstrom^2} per projection. 

We then performed mixed-state ptychographic reconstruction using the parameters in supplementary Table II, and averaged the resulting phase-contrast images. The resulting 2D class average is shown in the right panel of supplementary Fig. 4. We see that the ptychographic reconstruction from full-multislice simulated data is overall sharper than the experimental reconstruction, and the overall R-factor between experiment and mixed-state ptychographic reconstruction from multi-slice simulation is \SI{32}{\percent}. Possible reasons for the better quality of the reconstruction from simulated data, even including all partial coherence and dose effects, are imperfect detector quantum efficiency and modulation transfer function of the camera, which was not modeled in the simulations, and slight differences in the structure along the tube. Especially amorphous carbon wrapped around the tube, which was very helpful for alignment of the projections, causes resolution loss compared to the simulated structure.

Because of the better agreement of the linear model with the experiment, and the much lower computational complexity required for simulation, we performed subsequent tomographic reconstructions from model-generated projections with the linear model. We generate 28 projections with the linear forward model as above and reconstruct the volume with the same algorithm and alignment strategy as the experimental dataset, described in Section \ref{subsec:tomo}.
We then trace the atomic coordinates in the reconstructed volume with the same parameters as the experimental projections as described in Section \ref{sec:atom_tracing}, resulting in 575 traced atomic coordinates. The traced coordinates are then aligned to the experimental coordinates with the iterative closest point algorithm, and atoms obtained from simulation paired with the nearest traced atom from experiment. We then determine the displacement between experimental and simulated, retraced atoms. supplementary Fig. 6a shows a histogram of the RMS displacement of all 575 paired atoms, with a mean position error of \SI{17}{\pico\meter} and median position error of \SI{10}{\pico\meter}. supplementary Fig. 6b shows the spatial distribution of the displacement errors. It can be seen that Zr \& Te atoms close to the top and bottom of the nanotube, where the missing wedge artefacts are strongest, have the highest position error.

\subsection{Density functional theory calculations}
\label{sec:dft}
Previous first principles calculations on ZrTe$_5$ have emphasized the importance of accurately treating dispersive interactions to reproduce experimental structural parameters, with a comparison given in the SI, and the resulting electronic and topological properties \cite{Monserrat/Narayan:2019}. Because of this, we first performed structural optimizations on bulk ZrTe$_5$ to benchmark our choice of exchange-correlation functional with dispersive corrections. We find that the optB86b-vdW functional gives lattice parameters that are less than 1\% different than those measured by powder diffraction, a significant improvement over PBE which gives an error of over 9\% for the $b$ lattice parameter. (See further details in the SI).

Owing to the prohibitively large unit cell size of the full Zr-Te/CNT structure depicted in Fig.~\ref{fig:ZrTeModel}b which comprises over 5000 atoms for full \textit{ab initio} treatment, we instead perform calculations of the inner Zr-Te sandwich structure. To simulate the confining influence of the DW-CNT around the Zr-Te sandwich, we fix the outermost Te atoms during the structural optimization, while allowing the remaining Zr and Te positions to relax. We perform fixed-volume calculations for several values of the short \textit{a} lattice parameter, finding a minimum at 3.97 \AA{}, very close to the value of 3.95 \AA{} suggested from our diffraction analysis. As a test of the stability of the Zr$_{11}$Te$_{50}$ structure without the confining influence of the CNTs, we also performed a fixed-volume structural optimization allowing all atomic positions to relax \textit{without} including van-der-Waals corrections (i.e. using the PBE exchange correlation functional). With PBE only, we find the Zr$_{11}$Te$_{50}$ structure expands, which is primarily due to the separation between the structural units increasing and not changes in bond lengths or configurations, as we would expect without the dispersive corrections.

\noindent\textbf{Data and Code Availability:} 
The data and code used to produce the plots within this work will be released on the repository \texttt{Zenodo} upon publication of this preprint.

\noindent\textbf{Acknowledgments:} 
We would like to thank Gatan, Inc. as well as P. Denes, A. Minor, J. Ciston, J. Joseph, and I. Johnson who contributed to the development of the 4D Camera.

\noindent\textbf{Competing Interests:} 
The authors declare no competing interests.

\noindent\textbf{Funding Information:} 
PP and MCS are supported by the Strobe STC research center, Grant No. DMR 1548924.
CO acknowledges support from the Department of Energy Early Career Research Award program. Work at the Molecular Foundry was supported by the Office of Science, Office of Basic Energy Sciences, of the U.S. Department of Energy under Contract No. DE-AC02-05CH11231. 
SS, DP and AZ acknowledge support from the Director, Office of Science, Office of Basic Energy Sciences, Materials Sciences and Engineering Division, of the US Department of Energy under Contract No. DEAC02-05-CH11231: within the sp2-Bonded Materials Program (KC-2207) which provided for high resolution TEM imaging;  and within the van der Waals Heterostructures program (KCWF16) which provided for synthesis of the encapsulated material, and the National Science Foundation under Grant No. DMR-1807233 which provided for preliminary conventional TEM imaging.
This research used resources of the National Energy Research Scientific Computing Center (NERSC), a U.S. Department of Energy Office of Science User Facility located at Lawrence Berkeley National Laboratory, operated under Contract No. DE-AC02-05CH11231 using NERSC award ERCAP0020898 and ERCAP0020897.

\noindent\textbf{Author contributions:}
S.S. and D.P. prepared and screened CNT-encapsulated Zr-Te samples under supervision of A.Z. 
P.E. developed parts of the 4DCamera control and data management software.
P.M.P. carried out the 4D-STEM experiments, assisted by P.E. M.C.S. performed tilt axis alignment measurements.
P.M.P. implemented the data preprocessing codes and adapted the mixed-state ptychography reconstruction for 4DCamera data. P.M.P. and H.D. performed mixed-state ptychography reconstructions.
P.M.P. developed the tomographic reconstruction codes and performed tomographic reconstructions.
C.O. developed elliptical tube fitting codes, template matching, and unit cell extraction codes. P.M.P and C.O performed unit cell averaging and atom tracing. P.M.P. performed multi-slice simulations and precision estimation.  C.O, P.M.P., D.P. discussed and solved the atomic structure. S.G. performed DFT calculations, using input models created by C.O.
The study was conceived by P.M.P. under discussion with M.C.S, C.O. and S.S. P.M.P., C.O. M.C.S and S.G. wrote the manuscript with input from all authors.




\clearpage
\section*{Supplementary Material}
\setcounter{figure}{0}

\begin{table}[ht!]
\begin{tabular}{ l l}
 Microscope Voltage & \SI{80}{\kilo\eV} \\ 
 Electron gun & S-FEG\\  
 Source size (FWHM) & \SI{0.8}{\angstrom}\\  
 Cc & \SI{0.6}{\milli\meter}\\  
 Defocus spread (FWHM) & \SI{6}{\nano\meter}\\ 
 Convergence semi-angle & \SI{25}{\milli\radian}\\  
 Depth of focus & \SI{9}{\nano\meter}\\  
 Detector & 4D Camera @ \SI{87}{\kilo\hertz}   \\
 Detector pixel size & \SI{4.14e-3}{\angstrom^{-1}} / \SI{173.6}{\micro\radian}    \\
 Detector pixel size (binned) & \SI{0.0272}{\angstrom^{-1}} / \SI{1.136}{\milli\radian}   \\
 Detector outer angle & \SI{50}{\milli\radian}    \\
 Reconstruction pixel size & 0.397 Å\\
 Energy filter & No    \\
 Number of projections & 36    \\
 Tilt range & \SI{-53.9}{\deg}\\
 & \SI{65.2}{\deg}\\
 Total recorded patterns & \num{34560000}    \\
 Manually selected patterns & \num{8423259}    \\
 STEM step size & \SI{0.397}{\angstrom}    \\
 STEM dwell time & \SI{11.49}{\micro\second}    \\
 Probe current & \SI{40}{\pico\ampere}    \\
 Electron fluence accumulated & \SI{6.28e5}{\elementarycharge\per\angstrom^2}    \\
 Electron fluence per projection & \SI{1.72e4}{\elementarycharge\per\angstrom^2}    \\
 Avg. electrons per pattern & \num{2726}\\
 \end{tabular}
\caption{\label{tab:data_collection} Experimental parameters for PAET.}
\end{table}


\begin{figure*}[ht!]
    \includegraphics[width=1\textwidth]{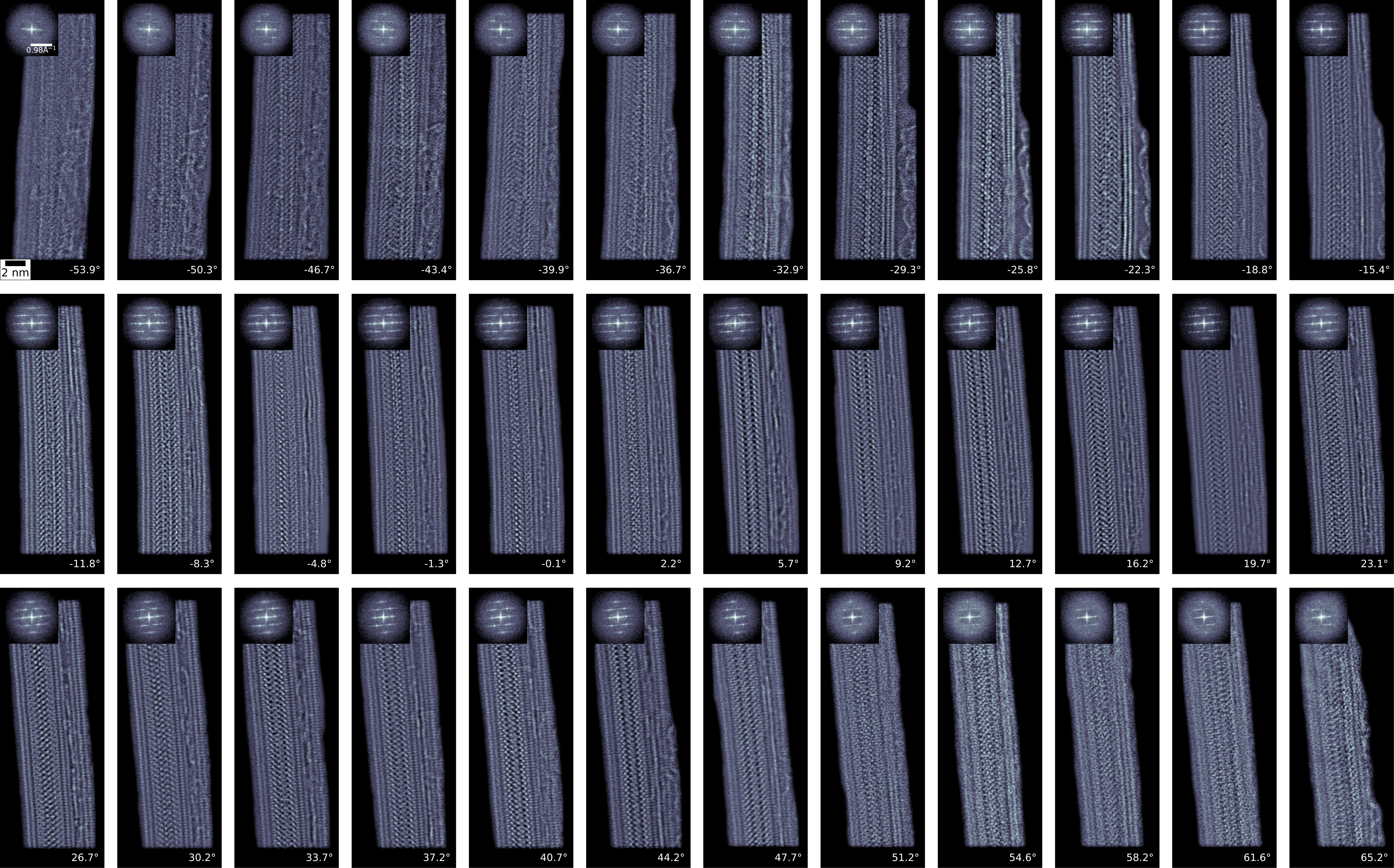}
    \caption{\label{fig:tilt_series} Ptychographic electron tomography tilt series of the the multi-walled carbon nanotube filled with $\mathrm{ZrTe_x}$. The 36 phase-contrast projection images with a tilt range from \num{-53.9} to \num{65.2} degrees (shown at bottom right of each panel) were measured with the 4D Camera and reconstructed using the LSQML algorithm. The total electron dose of the experiment is \SI{6.28e5}{\elementarycharge\per\angstrom^2}. The inset shows the power spectrum of each projection.}
\end{figure*}
\begin{figure*}[ht!]
    \includegraphics[width=1\textwidth]{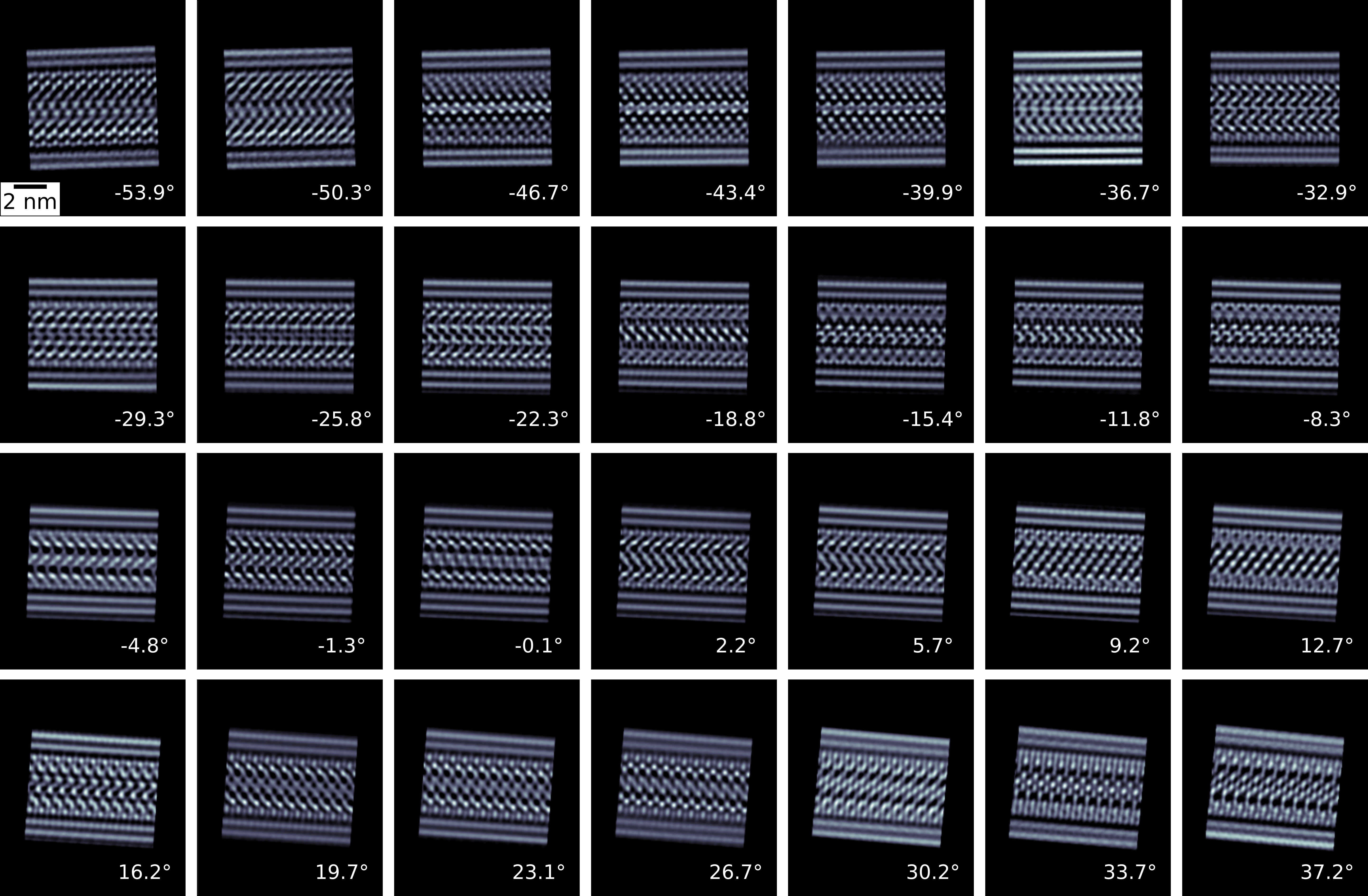}
    \caption{\label{fig:tilt_series_avg} Tilt series of unit-cell averaged ptychographic phase-contrast images from \SI{-53.9}{\degree} to \SI{37.2}{\degree}.}
\end{figure*}
\begin{table*}[ht!]
\begin{tabular}{ l l}
$\texttt{Nprobes}$ & 4\\ 
$\texttt{Nmodes}$ & 4\\  
$\texttt{method}$ & MLs\\  
$\texttt{Niter}$ & 115\\  
$\texttt{probe\_pos\_search}$ & 10\\ 
$\texttt{beta\_object}$ & 1\\  
$\texttt{beta\_probe}$ & 1\\
$\texttt{likelihood}$ & amplitude\\
$\texttt{grouping}$ & 200\\
$\texttt{apply\_subpix\_shift}$ & true\\
$\texttt{variable\_probe}$ & true\\
$\texttt{variable\_probe\_modes}$ & 1\\
$\texttt{variable\_intensity}$ & false\\
$\texttt{beta\_LSQ}$ & true\\
$\texttt{apply\_multimodal\_update}$ & false\\
$\texttt{delta\_p}$ & 0.1 \\
$\texttt{probe\_inertia}$ & 0.1\\
$\texttt{W}$ & 0\\
object initialization & constant phase\\
probe initialization & defocus from interactive \\
&single-sideband reconstruction \cite{Pelz_2021}\\
\end{tabular}
\caption{\label{tab:pramas} Reconstruction parameters for LSQML reconstruction.}
\end{table*}

\begin{figure*}[ht!]
    \includegraphics[width=1\textwidth]{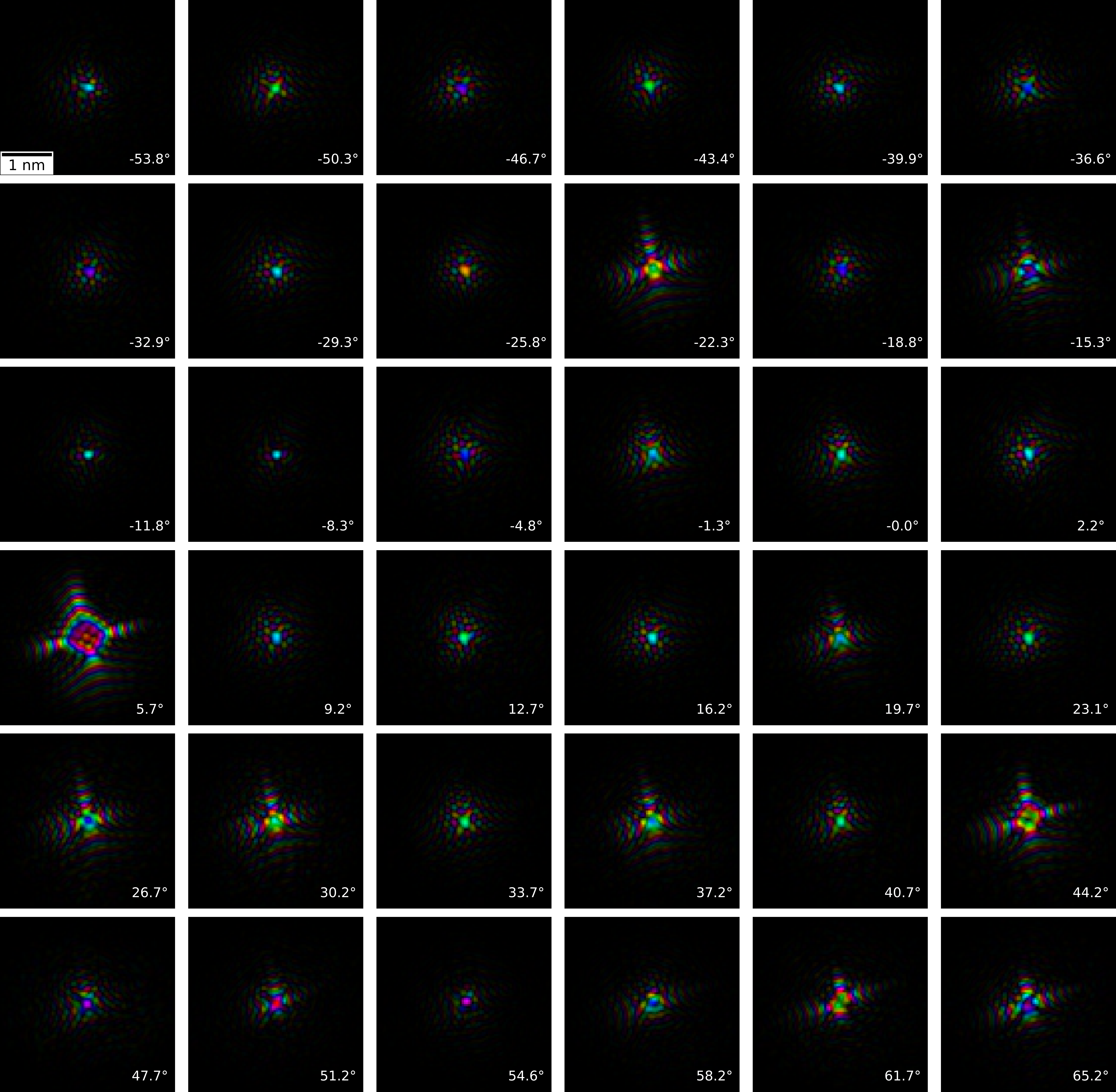}
    \caption{\label{fig:probes} Reconstructed probe wave function for each tilt. Amplitude is shown using color saturation, while phase of each pixel is given by the hue.}
\end{figure*}

\begin{figure*}[ht!]
    \includegraphics[width=0.7\textwidth]{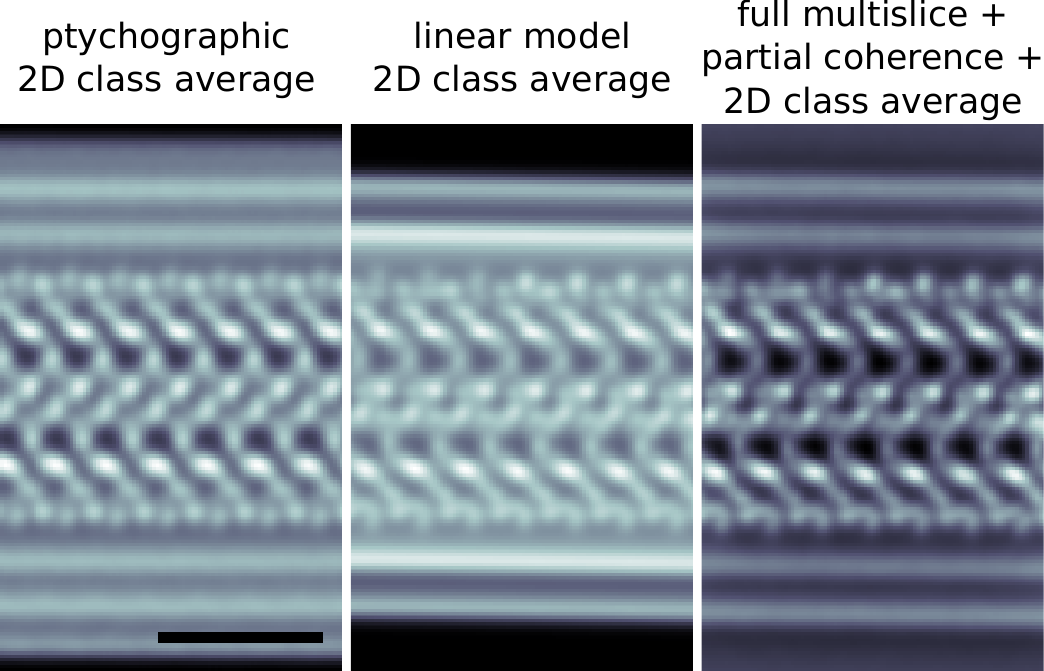}
    \caption{\label{fig:comparison_linear_model_multislice} Left panel: 2D class average obtained from experimental mixed-state ptychography reconstruction. Middle: 2D unit-cell average obtained from 12 atomic models where the carbon nanotube was shifted by multiples of \SI{17.75}{\pico\meter} along the tube direction. Right Panel: 2D unit-cell average obtained from 12 ptychographic reconstructions of 4D-STEM datasets simulated with the PRISM algorithm and partial spatial and temporal coherence effects accommodated. The carbon nanotube was shifted by multiples of 17.75 pm along the tube direction. Scale bar: \SI{2}{\nano\meter}}
\end{figure*}

\begin{figure*}[ht!]
    \includegraphics[width=1\textwidth]{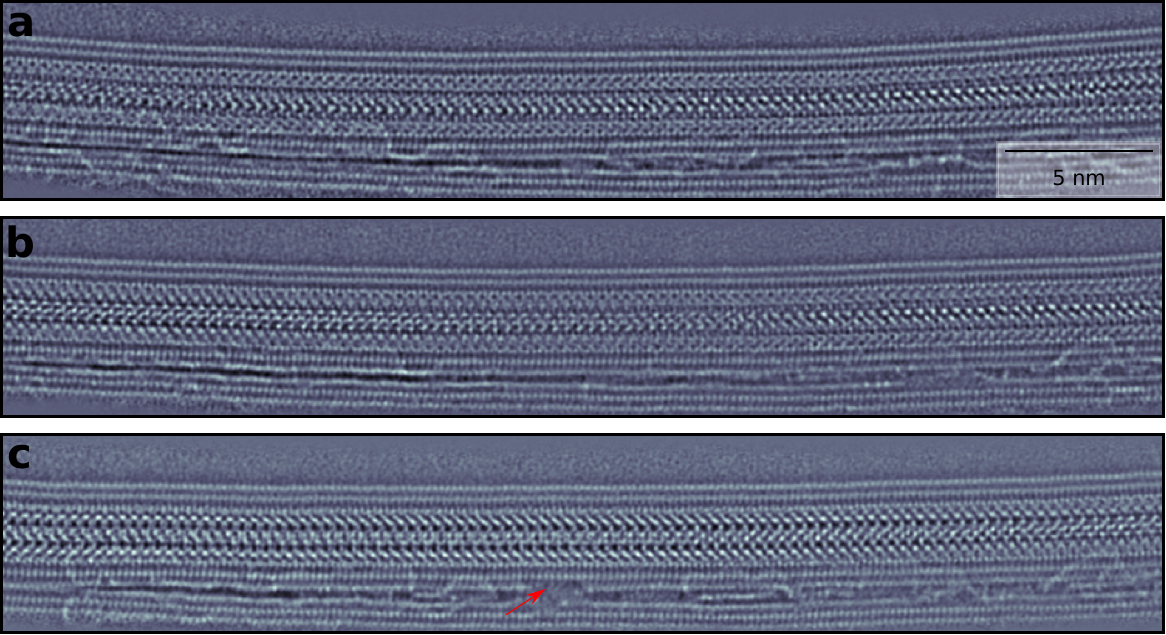}
    \caption{\label{fig:before_after} Reconstructed projections a) before, b) during and c) after the tilt series, close to \SI{0}{\degree} tilt. c) has a slightly different viewing angle due to tilt hysteresis. Only minor damage to the DW-CNT is visible after the tilt series is taken, indicated by the red arrow in c).}
\end{figure*}

\begin{figure*}[ht!]
    \includegraphics[width=1\textwidth]{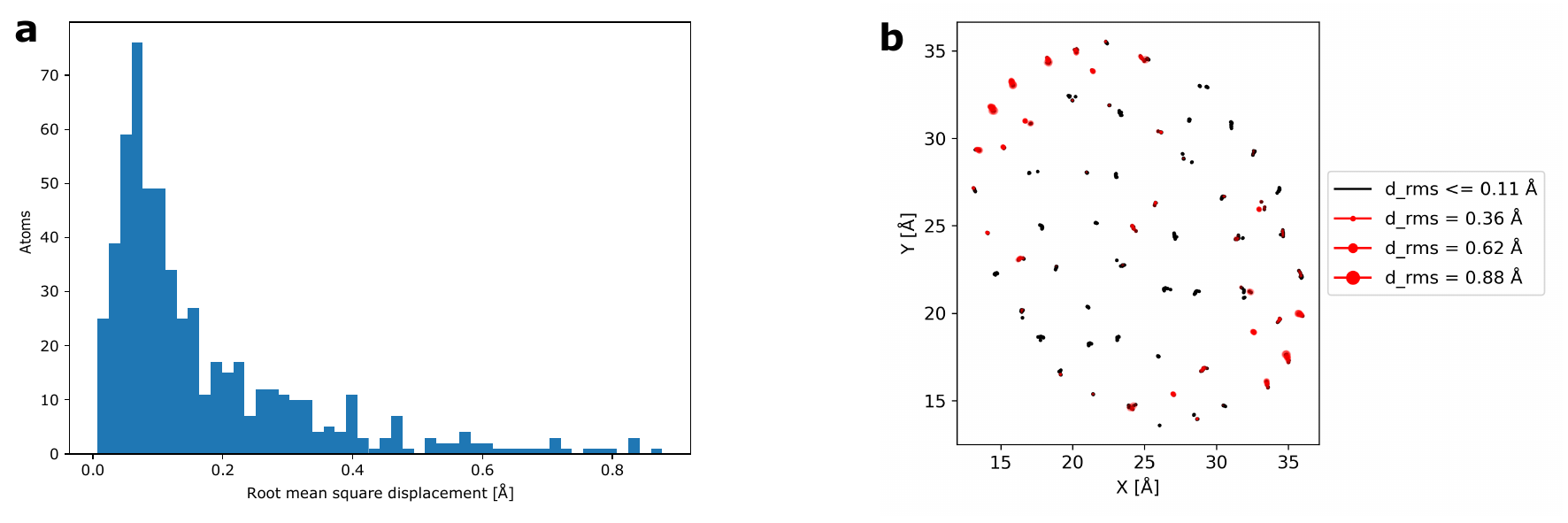}
    \caption{\label{fig:precision} a) Histogram of the distance between experimentally determined coordinates and coordinates obtained from traced coordinates from a simulated tilt series with the determined atomic model. Mean position error is \SI{17}{\pico\meter} and median position error is \SI{10}{\pico\meter}. b) Spatial distribution of the position errors of the Zr \& Te atoms, viewed along the nanotube.}
\end{figure*}

\begin{figure*}[ht!]
    \includegraphics[width=0.4\textwidth]{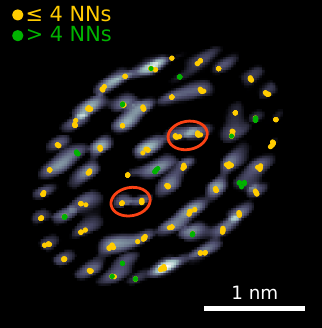}
    \caption{\label{fig:NNclassification} Projection of the core of the nanotube along the nanotube axis. Overlayed are the color-coded traced atomic positions depending on the number of the nearest neighbors (NNs). Atomic columns with predominantly less or equal 4 NNs were assigned as Te, while columns with predominantly greather than 4 NNs were asssigned Zr. The chemical species of the circled atoms was determined by choosing the lowest-energy configuration as determined by density functional theory.}
\end{figure*}

\newpage
\subsection*{Density Functional Theory calculation details}

All DFT calculations were carried out with the Vienna \textit{Ab initio} Simulation Package (\textsc{vasp}) \cite{Kresse1993,Kresse1994,Kresse1996} using projector augmented wave (PAW) pseudopotentials \cite{Blo,Kresse1999}.
Zr(4s, 4p, 5s, 4d) and Te(5s, 5p);  electrons were treated as valence , and the wave functions of the system were expanded in plane waves to an energy cutoff of 600 eV. Gamma-centered k-point grids of $14\times14 \times 4$, $12\times 12 \times  1$, and $12\times 12 \times 1$
were used for Brillouin Zone sampling of the bulk ZrTe$_5$, innercore Zr-Te tube, and central Zr-Te structure respectively. Dispersive corrections were accounted for using the optB86b-vdW of Klime\v{s} et al.~\cite{Klimes_et_al:2011} which gave calculated lattice parameters for bulk ZrTe$_5$ in the $Cmcm$ space group that are very close to those measured at 10 K \cite{Fjellvag_et_al:1986}. Spin-orbit coupling was not included in the structural optimizations as it was found to have very little effect on the structural parameters for bulk ZrTe$_5$, however it was included self-consistently in the electronic structure calculations. The electronic convergence criteria set to $10^{-7}$ eV and the force convergence criteria set to 0.002 eV / \AA{}.


Topological characterization using symmetry indicators and DFT calculated band structures was carried out using \textsc{symtopo} \cite{Symtopo}.
Previous work on ZrTe$_5$ discussed how the choice of exchange-correlational functional is essential for the accurate description of its structure, and hence electronic and topological properties \cite{Fjellvag_et_al:1986, Nair_et_al:2018, Monserrat/Narayan:2019}. We find that the optB86b-vdW of Klime\v{s} et al.~ \cite{Klimes_et_al:2011} gives lattice parameters extremely close to those measured in experiment \cite{Fjellvag_et_al:1986}.

\clearpage

\bibliography{main}

\end{document}